\documentclass[twocolumn,superscriptaddress,preprintnumbers,amsmath,amssymb,prl]{revtex4}
\usepackage{graphicx}
\usepackage{dcolumn,xcolor}
\usepackage[normalem]{ulem}
\usepackage{amsmath} 
\usepackage{amssymb}
\usepackage{amsfonts}
\usepackage{bm}
\usepackage[latin1]{inputenc}
\usepackage[dvips]{epsfig}
\usepackage{hyperref}
\usepackage{color}

\usepackage{color}
\usepackage{tikz}
\usetikzlibrary{shapes}

\newcommand{\circleNUM}{\raisebox{0.5pt}{\tikz{\node[draw,scale=0.5,circle,fill=blue!65!green](){};}}}

\newcommand{\invertedtriangleNUM}{\raisebox{0pt}{\tikz{\node[draw,scale=0.35,regular polygon, regular polygon sides=3,fill=white!10!green,rotate=180](){};}}}

\newcommand{\triangleNUM}{\raisebox{0.5pt}{\tikz{\node[draw,scale=0.35,regular polygon, regular polygon sides=3,fill=blue!40!green,rotate=0](){};}}}

\newcommand{\squareNUM}{\raisebox{0.5pt}{\tikz{\node[draw,scale=0.5,regular polygon, regular polygon sides=4,fill=white!1!blue](){};}}}

\newcommand{\circleEXP}{\raisebox{0.5pt}{\tikz{\node[draw,scale=0.5,circle,fill=black!25!brown](){};}}}

\newcommand{\invertedtriangleEXP}{\raisebox{0pt}{\tikz{\node[draw,scale=0.35,regular polygon, regular polygon sides=3,fill=white!50!orange,rotate=180](){};}}}

\newcommand{\triangleEXP}{\raisebox{0.5pt}{\tikz{\node[draw,scale=0.35,regular polygon, regular polygon sides=3,fill=red!30!brown,rotate=0](){};}}}

\newcommand{\squareEXP}{\raisebox{0.5pt}{\tikz{\node[draw,scale=0.5,regular polygon, regular polygon sides=4,fill=black!70!brown](){};}}}

\definecolor{fig1_1}{rgb}{1.000000000000000 ,  0.781200000000000  , 0.497500000000000}

\newcommand{\gp}{\dot\gamma}
\newcommand{\gm}{\gamma_\text{M}}

\newcommand{\sm}{\sigma_\text{M}}
\newcommand{\sy}{\sigma_\text{y}}
\newcommand{\Gp}{\dot\Gamma}
\newcommand{\Gm}{\Gamma_\text{M}}
\newcommand{\Gy}{\Gamma_1}
\newcommand{\Sm}{\Sigma_\text{M}}

\newcommand{\tit}{\tilde{t}}

\newcommand{\titm}{\tit_\text{M}}
\newcommand{\titone}{\tit_\text{1}}

\newcommand{\seb}[1]{\textcolor{black}{#1}}
\newcommand{\MS}[1]{\textcolor{black}{#1}}

\newcommand{\tib}[1]{\textcolor{black}{ #1}}

\begin{document}

\title{Stress Overshoots in Simple Yield Stress Fluids}

  \author{Roberto Benzi}
 \affiliation{Dipartimento di Fisica, Universit\`a di Roma ``Tor Vergata" and INFN, Via della Ricerca Scientifica, 1-00133 Roma, Italy\looseness=-1}
 \author{Thibaut Divoux}
\affiliation{Univ Lyon, Ens de Lyon, Univ Claude Bernard, CNRS, Laboratoire de Physique, F-69342 Lyon, France\looseness=-1}
\author{Catherine Barentin}
  \affiliation{Universit\'e de Lyon, Universit\'e Claude Bernard Lyon 1, CNRS, Institut Lumi\`ere Mati\`ere, F-69622 Villeurbanne, France\looseness=-1}
 \author{S\'ebastien Manneville}
\affiliation{Univ Lyon, Ens de Lyon, Univ Claude Bernard, CNRS, Laboratoire de Physique, F-69342 Lyon, France\looseness=-1}
\author{Mauro Sbragaglia}
\affiliation{Dipartimento di Fisica, Universit\`a di Roma ``Tor Vergata" and INFN, Via della Ricerca Scientifica, 1-00133 Roma, Italy\looseness=-1}
\author{Federico Toschi}
\affiliation{Department of Applied Physics, Eindhoven University of Technology, P.O. Box 513, 5600 MB Eindhoven, The Netherlands and CNR-IAC, Rome, Italy.}

\date{\today}

\begin{abstract}
Soft glassy materials such as mayonnaise, wet clays, or dense microgels display a solid-to-liquid transition under external shear. Such a shear-induced transition is often associated with a non-monotonic stress response, in the form of a stress maximum referred to as ``stress overshoot". This ubiquitous phenomenon is characterized by the coordinates of the maximum in terms of stress $\sm$ and strain $\gm$ that both increase as weak power laws of the applied shear rate. Here we rationalize such power-law scalings using a continuum model that predicts two different regimes in the limit of low and high applied shear rates. The corresponding exponents are directly linked to the steady-state rheology and are both associated with the nucleation and growth dynamics of a fluidized region. Our work offers a consistent framework for predicting the transient response of soft glassy materials upon start-up of shear from the local flow behavior to the global rheological observables.
\end{abstract}

\maketitle

\begin{figure}[tb]
    \centering
    \includegraphics[width=0.9\columnwidth]{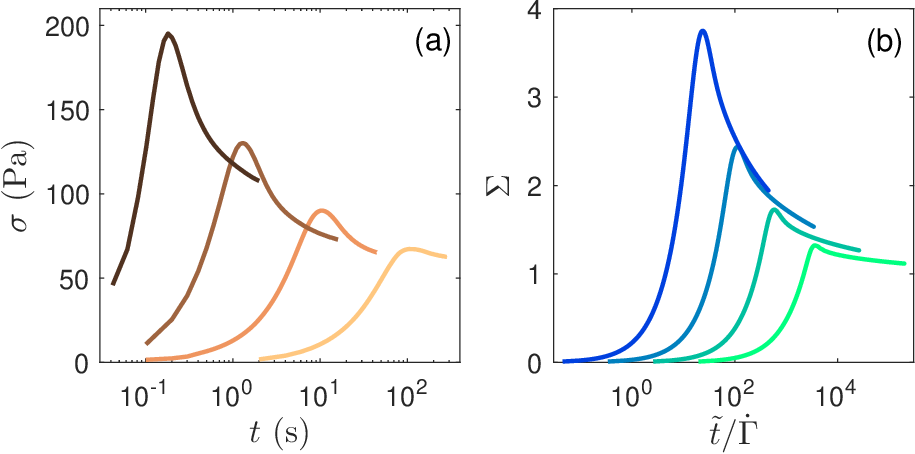}
        \caption{Phenomenology of the stress overshoot during start-up of shear. Stress responses (a)~$\sigma$ as a function of time $t$ in experiments \seb{performed on a 1~\% wt.~Carbopol microgel} (adapted from Fig.~3 in Ref.~\cite{Divoux:2011}) under imposed shear rates $\gp=10$, 1, 0.1, and 0.01~s$^{-1}$ from left to right and (b)~the present fluidity model for $\Gp=0.22$, $2.9\times 10^{-2}$, $3.8\times 10^{-3}$ and $5.1\times 10^{-4}$ with a fixed value of $\tau=1$ (see text for details).
        }
    \label{fig:overshoots}
\end{figure}

\textit{Introduction.-} From dense suspensions and gels to metallic alloys and composites, numerous materials display a non-monotonic stress response under external shear. For a given applied shear rate $\gp$, the stress increases up to a maximum $\sm$ reached at a strain $\gm$ before decreasing towards its steady-state value, while the sample yields (see Fig.~\ref{fig:overshoots}). This sequence, also referred to as the ``stress overshoot", is a complex process to model as it depends on the applied shear rate as well as the details of the sample microstructure through the sample age, its thermal and shear history, etc.  \cite{Rodney:2011,Siebenburger:2012a,Dimitriou:2014,Bonn:2017,Joshi:2018,Ozawa:2018}. 

Soft Glassy Materials (SGMs) encompass soft amorphous systems such as gels and glasses. These materials are characterized by a yield stress $\sy$ below which the sample responds as a solid, and above which it flows like a liquid \cite{Bonn:2017}. Under external shear, most SGMs display a stress overshoot, which results from the rearrangement of the sample microstructure. The stress peak is correlated to the maximum structural anisotropy \cite{Mohraz:2005,Koumakis:2012}, while the subsequent stress relaxation is \tib{dominated by nonaffine displacements, and} associated with either \tib{cage breaking and} super-diffusive motion of particles in the case of glasses \cite{Zausch:2008,Koumakis:2012,Laurati:2017}, or strand failure in the case of gels \cite{Whittle:1997,Park:2013,Santos:2013}. Concomitantly to the stress relaxation, the sample may either flow homogeneously or show the formation of transient or steady-state shear bands, or even fracture \cite{Magnin:1990,Persello:1994,Divoux:2010,Moorcroft:2011,Amann:2013}.

Despite such complexity, the amplitude $\sm$ of the stress overshoot consistently increases as a power law of $\gp$, with an exponent that varies from 0.1 to 0.5 as reported in experiments on gels and repulsive glasses \cite{Derec:2003,Carrier:2009,Divoux:2011,Koumakis:2011,Amann:2013}. Stress overshoots are well reproduced by various \tib{theoretical} approaches such as Brownian or molecular dynamics simulations, micromechanical modelling and Mode Coupling Theory, which have provided valuable insights on the microscopic scenario associated with the overshoot \cite{Utz:2000,Amann:2013,Colombo:2014,Zaccone:2014,Khabaz:2021}. However, the functional form $\sm(\gp)$ inferred from computations is most often either logarithmic \cite{Rottler:2003,Varnik:2004,Rottler:2005,Shrivastav:2016} in contradiction with experimental results, or a power law with an exponent 0.5 \cite{Whittle:1997,Park:2013,Johnson:2018}, which does not reflect the broad range of exponents reported in the literature.
A noticeable exception is the seminal version of the fluidity model, which yields a power-law scaling, with exponents lower than 0.5 \cite{Derec:2003}. However, to date, there is no consistent theoretical framework offering a rationale for the multiplicity of power-law exponents reported for stress overshoots in SGMs.

In this Letter, we tackle the case of ``simple'' Yield Stress Fluids (YSFs), a subclass of SGMs whose steady-state flow is homogeneous and described by a Herschel-Bulkley rheology \cite{Bonn:2017}. We use a model first introduced in Ref.~\cite{Bocquet:2009} \seb{based on a {\it fluidity} parameter,} and successfully extended to capture the spatially-resolved yielding scenario of SGMs \cite{Benzi:2016,Benzi:2019}, to rationalize the effect of shear on the coordinates ($\gm$, $\sm$) of the stress overshoot. We show that the relevant variable to quantify the magnitude of the stress overshoot is \seb{$\sm/\sy-1$, and that this normalized parameter displays two asymptotic power-law regimes as a function of the applied shear rate, namely}, a diffusive regime for low shear rates and an asymptotic scaling at large shear rates. In both cases, the value of the exponent is set by the power-law constitutive behavior in steady-state flow. Finally, our approach allows us to account not only for the shear dependence of the stress overshoots but also for the local flow behavior upon start-up of shear.

\textit{Fluidity model.-} We consider a simple YSF whose steady-state rheology follows the Herschel-Bulkley (HB) model, which reads $\Sigma=1+\Gp^n$ in dimensionless units, where $\Sigma=\sigma/\sy$ is the shear stress normalized by the yield stress, and $\Gp=\gp/(\sy/A)^{1/n}$ is the shear rate normalized by the natural frequency for the HB model, \seb{$\sigma=\sy+A\gp^n$}, with $n$ the HB exponent and $A$ the consistency index. The fluid is sheared between two walls, separated by a distance $L$ and its dynamics is encoded in the local dimensionless fluidity $f(y)$, where $y$ is the spatial coordinate along the velocity gradient direction. \MS{As originally introduced in Ref.~\cite{Bocquet:2009}, the fluidity is a dynamical coarse-graining parameter related to the rate of plastic events. More intuitively, one can consider the fluidity as the inverse of the viscosity.} We also define the rescaled time $\tit \equiv \gp t$, which corresponds to the physical strain. As discussed in Ref.~\cite{Benzi:2019}, the fluid rheological response is well described by the following equation for the fluidity:
\begin{equation} \label{eq1}
\frac{\partial f}{\partial \tit}= f[\xi^2\Delta f+m f-f^{3/2}]\,,
\end{equation}
where $\xi$ is the so-called cooperativity length \seb{and relates to the extension of the region that is impacted by a neighboring plastic rearrangement} \cite{Goyon:2008,Bocquet:2009,Goyon:2010,Geraud:2013,Geraud:2017}, and $m=m(\Sigma)$ with $m^2=(\Sigma-1)^{1/n}/\Sigma$ for $\Sigma \geq 1$ and $m=0$ for $\Sigma < 1$. \seb{The latter parameter $m$ essentially conveys the information about the underlying steady-state HB rheology, as $f=m^2$ corresponds to the stationary homogeneous solution of Eq.~\eqref{eq1}.} Moreover, we assume a simple plane shear flow and that $\Sigma$ is spatially homogeneous and only depends on $t$. To model the response to an imposed shear rate $\Gp$, Eq.~\eqref{eq1} is coupled to the following evolution equation for the stress based on a Maxwell model:
\begin{equation} \label{eq2}
\frac{\text{d} \Sigma}{\text{d} \tit}=\frac{G_0}{\sy}\left (1-\frac{\langle f \rangle \Sigma}{\Gp} \right)\,,
\end{equation}
where $G_0$ is the elastic modulus, and $\langle f \rangle$ is the spatial average of the fluidity, which is a function of time. We have shown that this approach successfully captures the the long-time evolution of SGMs towards steady state \cite{Benzi:2016,Benzi:2019}. Here we explore the short-time response of this model during shear start-up. As a generic case, we solve Eqs.~\eqref{eq1} and \eqref{eq2} with $n=1/2$, for fixed values of $\Gp$ ranging between $10^{-4}$ and $10^2$ with $\xi/L=0.04$ and assuming $ f (y,0) =  10^{-4} \ll 1$ for the initial solid-like state and $ f(0,\tit) = m^2(\Sigma(\tit))$ and $\partial_{y}  f (L, \tit)=0$ for boundary conditions at the two different walls. In this framework, we explore the behavior of a simple YSF with respect to two parameters, i.e., the imposed shear rate $\Gp$ and the dimensionless relaxation time $\tau=\sy/G_0$. 

\begin{figure}
    \centering
    \includegraphics[width=0.85\columnwidth]{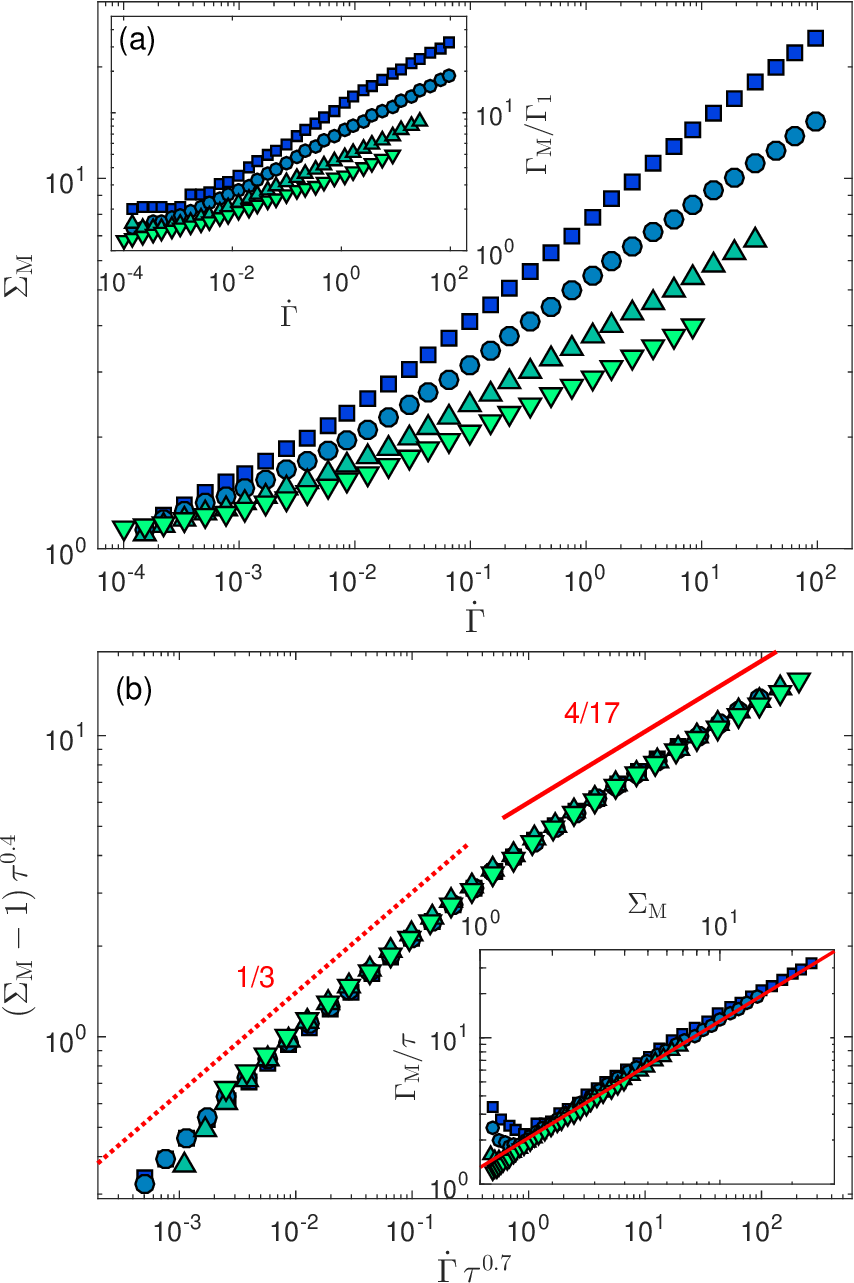}
    \caption{Analysis of the stress overshoots predicted by the fluidity model for an HB exponent $n=1/2$. (a)~Stress maximum $\Sm$ vs shear rate $\Gp$. Inset: strain at maximum $\Gm$ normalized by the strain $\Gy$ that corresponds to $\Sigma=1$ and plotted against $\Gp$. Colored symbols refer to different values of $\tau=0.1$~(\protect \squareNUM), 1~(\protect \circleNUM), 10~(\protect \triangleNUM), and 100~(\protect \invertedtriangleNUM).
    (b)~Rescaled stress maximum $(\Sm-1)\tau^{0.4}$ vs $\Gp\tau^{0.7}$. Inset: $\Gm/\tau$ vs $\Sm$. The red line is $\Gm/\tau=1.3\,\Sm$.}
    \label{fig:fluidity}
\end{figure}

\textit{Theoretical scalings.-} As illustrated in Fig.~\ref{fig:overshoots}(b) for four values of $\Gp$, the model predicts stress overshoots very similar to those reported in experiments, \seb{e.g., on carbopol microgels [Fig.~\ref{fig:overshoots}(a)].} More generally, extracting \seb{the stress maximum} $\Sm(\Gp)$ and the corresponding strain $\Gm(\Gp)=\titm$ for values of $\tau$ spanning three orders of magnitude, we find that \seb{$\Sm$ grows faster with $\Gp$ as $\tau$ decreases, i.e., when the elastic modulus $G_0$ increases relative to the yield stress $\sy$} [Fig.~\ref{fig:fluidity}(a)]. As a central result of this Letter, we show that the entire data set can be rescaled onto the master curve of Fig.~\ref{fig:fluidity}(b), which is composed of two power-law asymptotic limits, namely $(\Sm-1) \sim \Gp^{1/3}$ for $\Gp \ll 1$, and $(\Sm-1) \sim \Gp^{4/17}$ for $\Gp \gg 1$. These two limits are justified analytically \seb{in detail in the companion paper \cite{companion} and can be understood qualitatively as the signature of two different dynamical regimes for the nucleation and growth of a shear band of size $\ell_b(\tit)$ at the moving wall. Indeed, upon shear start-up, the initial fluidity remains negligible and the stress $\Sigma(\tit)$ grows roughly linearly} up to $\Sigma(\titm)=\Sm$ where the l.h.s.~of Eq.~\eqref{eq2} must be zero, yielding $\Gp= \langle f \rangle(\titm)\Sm$.
Since the fluidity $\langle f \rangle$ is dominated by the fluidity in the \seb{shear band}, we may approximate $f$ with the value of $m^2(\tit)$ for any applied shear rate, yielding $\langle f \rangle(\tit) \sim \ell_b(\tit) \, m^2(\tit)/L$, \seb{with $m^2(\tit)\sim\Sigma(\tit)\sim\Sigma(\tit)-1$ under the assumption that $\Sigma(\tit) \gg 1$.}

In the limit of large shear rates, the fluidity grows from the moving wall, triggering the formation of a fluidized front. \seb{Scaling arguments show that the characteristic length and time in the system are $\xi/\sqrt{m}$ and $m^{-3}$ respectively~\cite{Benzi:2019,companion}. Hence, the shear band is expected to move with a velocity $\text{d} \ell_b/\text{d} \tit \sim m^3 \xi/\sqrt{m} = \xi m^{5/2}(\tit)$. Using $m(\tit)\sim[\Sigma(\tit)-1]^{1/2}$ and integrating over time} yields
$\ell_b(\tit) \sim \xi \tau [ (\tit-\titone)/\tau]^{9/4}$, where $\titone$ is the time at which the stress equals the yield stress, i.e., $\Sigma(\titone)=1$. Finally, combining the expression of $m(\tit)$ and $\ell_b(\tit)$ \seb{at $\tit=\titm$ and using $\Sm-1\sim \titm-\titone$} leads to an expression for $\langle f \rangle(\titm)$ \seb{and to the following} asymptotic scaling:
\begin{equation}
\Sm-1 \sim \left(\frac{\dot \Gamma}{\xi \tau}\right)^{4/17} \seb{~\text{for}~\Gp \gg 1}.
\label{eq:lowshear}
\end{equation}

\begin{figure}[t!]
    \centering
   \includegraphics[width=0.9\columnwidth]{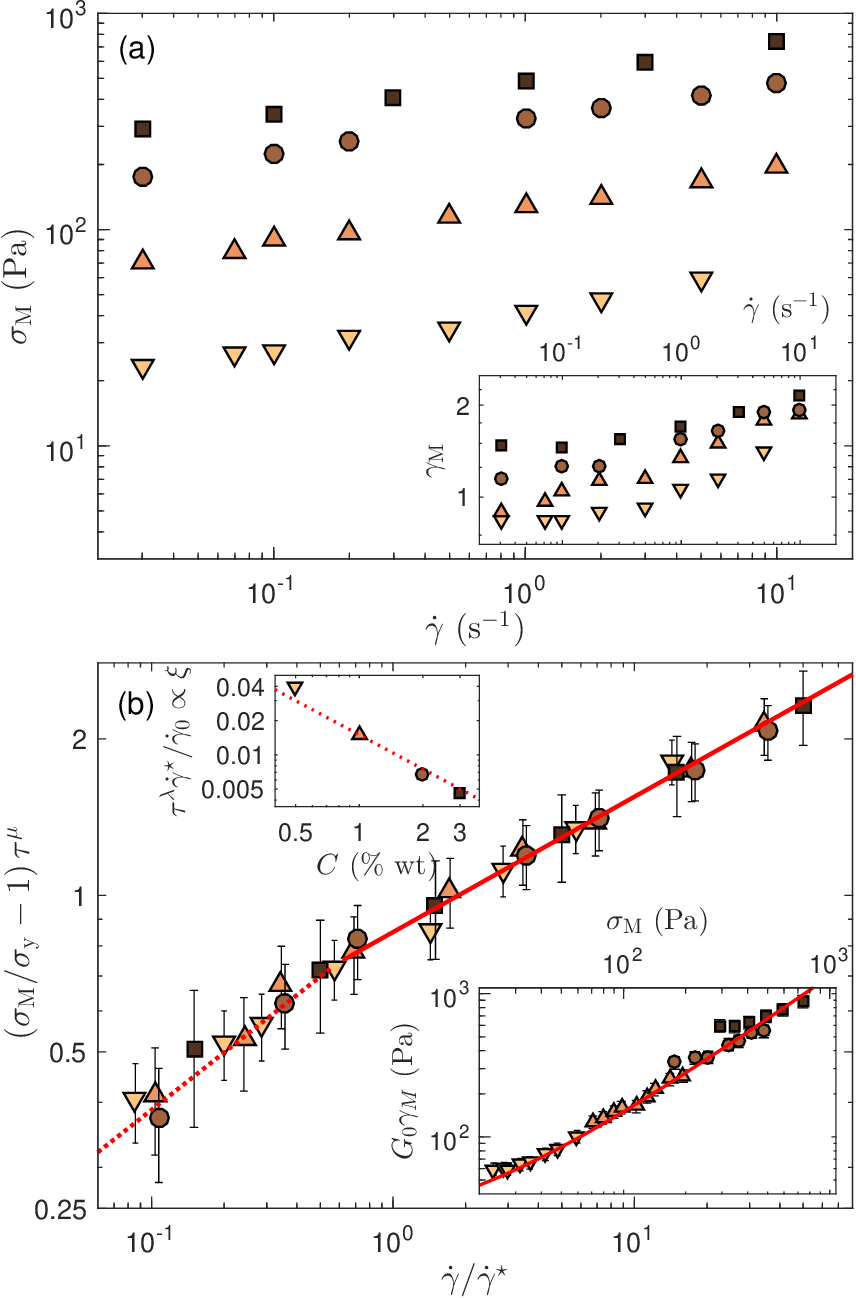}
    \caption{Analysis of the stress overshoots recorded in experiments on Carbopol microgels with different concentrations $C=0.5\%$~(\protect \invertedtriangleEXP), 1\%~(\protect \triangleEXP), 2\%~(\protect \circleEXP), and 3\%~wt~(\protect \squareEXP). (a)~Stress maximum $\sm$ as a function the applied shear rate $\gp$. Inset: corresponding strain $\gm$ vs $\gp$. Data from Ref.~\cite{Divoux:2011}. \seb{The experimental uncertainty on these raw data is smaller than the symbol size.} (b)~Rescaled stress maximum $(\sm/\sy-1)\tau^{\mu}$ vs \seb{$\gp/\gp^\star$}, where $\tau=\sy/G_0$, with $G_0$ the elastic modulus of the microgel. The red solid line is a power law with exponent $\bar{\alpha}=\langle 4n/(9-n)\rangle=0.27\pm 0.01$ inferred from the fluidity model in the asymptotic regime and averaged over the various samples. The red dotted line shows the scaling predicted in the diffusive regime with exponent $\bar{\beta}=\langle 2n/3\rangle=0.38\pm 0.01$. Lower inset: $G_0\gm$ vs $\sm$. The red line is $G_0\gm=1.3\,\sm+20$. See Supplemental Material Table S1 for the values of $\sy$, $n$, and $G_0$ used in the rescaling. \seb{Error bars account for the experimental uncertainty on $\sy$ and $G_0$. Upper inset: rescaling factor $\gp^\star$ used for the shear rate as a function of $C$ and normalized by $\gp_0\tau^{-\lambda}$ with $\gp_0=(\sy/A)^{1/n}$ so as to provide a quantity that is predicted by the theory to be proportional to the cooperativity length $\xi$. The red dotted line shows that this quantity scales roughly as $1/C$. The exponents $\lambda$ and $\mu$ are inferred from the HB exponent $n$, which depends weakly on the Carbopol concentration \cite{note_general_HB}.}}
    \label{fig:carbopol}
\end{figure}

In the limit of low shear rates, the system reorganises in the vicinity of the moving wall without any propagating front solution. Plastic activity rather occurs via diffusion effects, which are peculiar to the fluidity equation [Eq.~\eqref{eq1}], so that the size of the shear band follows a diffusive growth $\ell_b(\tit) \sim \xi m(\tit) \left(\tit-\titone\right)^{1/2}$. Following the same steps as in the high shear rate limit, we get:
\begin{equation}
\Sm-1 \sim \left(\frac{\Gp}{\xi \tau^{1/2}}\right)^{1/3} \seb{~\text{for}~\Gp \ll 1}.
\label{eq:highshear}
\end{equation}
As shown in the inset of Fig.~\ref{fig:fluidity}(b), the strain $\Gm$ is simply proportional to $\Sm$ so that the two asymptotic scalings also hold for $\Gm/\Gy-1$. Finally, combining Eqs.~\eqref{eq:lowshear} and \eqref{eq:highshear}, we expect the transition between the two behaviors to occur at $\dot\Gamma^{\star}\sim \seb{\xi} \tau^{-0.7}$ and $\Sigma^{\star}_\text{M}-1\sim \tau^{-0.4}$. As shown in Fig.~\ref{fig:fluidity}(b), the rescaled data $(\Sm-1)\tau^{0.4}$ as function of $\dot\Gamma\tau^{0.7}$ indeed nicely collapse onto the predicted master curve over the whole range of studied shear rates. As detailed in the companion paper~\cite{companion}, the above approach can be generalized to any value $n$ of the HB exponent, leading to scalings with exponents $\alpha=4n/(9-n)$ at large shear rates and $\beta=2n/3$ at small shear rates instead of 4/17 and 1/3 in Eqs.~\eqref{eq:lowshear} and \eqref{eq:highshear} respectively.

\textit{Discussion.-} Let us now compare our theoretical findings against experimental data. We revisit the shear start-up experiments of Ref.~\cite{Divoux:2011} performed on Carbopol microgels for concentrations ranging between 0.5\% and 3\%~wt.~in a parallel-plate geometry connected to a stress-controlled rheometer (see also Supplemental Material for details). Such a simple YSF displays a stress overshoot upon shear start-up [Fig.~\ref{fig:overshoots}(a)]. As reported in Fig.~\ref{fig:carbopol}(a), both the stress maximum $\sm$ and \seb{the corresponding strain} $\gm$ increase weakly with the applied shear rate $\dot\gamma$. When considering $\sm/\sy-1$, the experimental data for the stress can be further rescaled into a single master curve spanning more than two decades [Fig.~\ref{fig:carbopol}(b)], which displays two asymptotic scalings in excellent agreement with the two exponents $\alpha$ and $\beta$ derived from the fluidity model for an arbitrary value of $n$. Moreover, when multiplied by the elastic modulus $G_0$, the strain $\gm$ collapses onto a single affine law of $\sm$ with the same prefactor as $\Gm/\tau$ in the theory [lower inset of Fig.~\ref{fig:carbopol}(b)]. Therefore, our theoretical approach nicely captures the early stage response of this SGM to shear, as well as its subsequent fluidization \cite{Benzi:2019}.

\begin{figure}[t!]
    \centering
    \includegraphics[width=0.9\columnwidth]{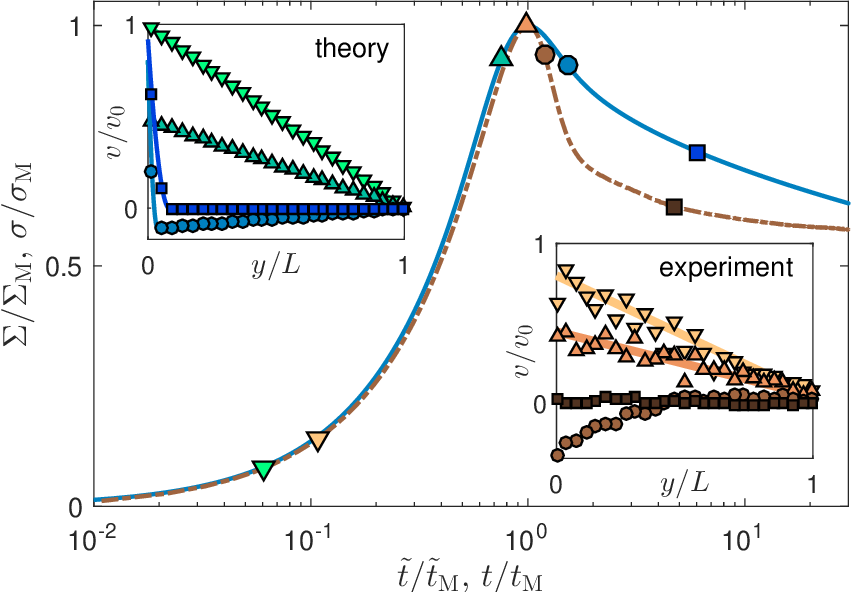}
    \caption{Rescaled stress response $\Sigma/\Sm$ vs $\tit/\titm$ from the fluidity model (blue solid line, $\Gp=0.029$ and $\tau=1$) and $\sigma/\sm$ vs $t/t_\text{M}$ from experiments on a carbopol microgel (brown dash-dotted line, $\gp=0.1$~s$^{-1}$, $C=1$\%~wt, adapted from Fig.~4 in Ref.~\cite{Divoux:2011}). $\titm$ ($t_\text{M}$ resp.) corresponds to the strain (time resp.) at which the maximum stress $\Sm$ ($\sm$ resp.) is reached. Insets: velocity profiles $v$ normalized by the velocity of the moving plate $v_0$ as a function of the distance $y$ to the moving plate normalized by the gap size $L$ and taken at the various times indicated by the symbols in the main graph. Lines on the experimental profiles are guides for the eye.}
    \label{fig4}
\end{figure}

Beyond the quantitative prediction of the locus of the stress maximum, our theoretical approach allows us to compute the local velocity profiles during shear start-up. Figure~\ref{fig4} shows the velocity profiles at various times along the stress response of the material predicted for $\Gp=0.029$ and $\tau=1$. The velocity profile is linear during the initial growth of the stress, which is indicative of affine displacement during the initial stage. Around the stress maximum, the fluidity at the moving wall becomes sufficiently large that the shear rate in the bulk decreases, leading to an elastic recoil after which the velocity profile flattens out. These results are in excellent agreement with the experimental observations on carbopol microgels, in which the formation of a thin lubrication layer at the wall leads to a fast recoil followed by a total wall slip regime (lower inset in Fig.~\ref{fig4})~\cite{Divoux:2011}. 

Our theoretical approach provides the following rationale for the observed phenomenology. When shear is switched on, the fluidity at the wall and the thickness $\ell_b$ of the shear band are small. At short times, and thus for small $\ell_b$, the second term on the r.h.s of Eq.~\eqref{eq2} does not play any significant role and the stress grows in time almost linearly. Around the stress maximum, the system enters a different dynamical regime where ${\rm d}\Sigma/{\rm d}\tit\simeq 0$, i.e., the instantaneous value of the stress is balanced by the effective shear rate within the shear band $\Gp L/\ell_b$~\cite{companion}. The clear-cut separation between these two different dynamical behaviours allows us to provide a distinctive theoretical prediction for the scaling of the stress maximum as a function of the shear rate.
Finally, note that the narrow fluidized band near the wall eventually grows into a transient shear band, whose dynamics and lifespan have been extensively discussed in Ref.~\cite{Benzi:2019}.   
 
To conclude, the present fluidity model encompassing non-local effects provides a comprehensive framework for describing the stress overshoot that goes along with the start-up of shear in simple YSFs. Our approach shows that the relevant observables are $\Sm-1$ and $\Gm-1$ instead of the raw values of the stress maximum coordinates. In that framework, our model yields a quantitative prediction for the rate dependence of the overshoot in the form of two power-law scalings in the limits of low and high shear rates, which may apply to a vast amount of data from the literature --see companion paper \cite{companion} for additional comparisons with previous experimental and numerical results \cite{Varnik:2004,Amann:2013,Carrier:2009,Fernandes:2017}. Non-local effects play a key role in the predicted scalings by \seb{governing} the growth of the shear band nucleated in the vicinity of the moving wall: \seb{depending on the strain rate, fluidization is driven either by diffusive dynamics or by front propagation}. This scenario provides an alternative to existing descriptions of the stress overshoot in terms of local rearrangements and cage dynamics \cite{Koumakis:2012,Khabaz:2021,Varnik:2004} and to continuum viscoelastic models based on recoverable strain measurements or mean-field elastoplastic models \cite{Ozawa:2018,Donley:2020,Singh:2021,Kamani:2021}. As also emphasized in the companion paper~\cite{companion}, SGMs forming permanent shear bands can be captured within a generalized version of our fluidity model. We show that such a generalization does not affect the scaling properties of the overshoot and that further including long-range correlations into the model and playing with boundary conditions can account for avalanche-like effects as well as brittle-like vs ductile-like response past the overshoot. In that respect, our results should set a basis for predicting shear start-up flow in a wide variety of SGMs.

\begin{acknowledgments}
The authors thank David Tamarii for help with the experiments. This research was supported in part by the National Science Foundation under Grant No. NSF PHY-1748958 through the KITP program on the Physics of Dense Suspensions. \MS{This work received funding from the European Research Council (ERC) under the European Union's Horizon 2020 research and innovation
programme (grant agreement No 882340).}
\end{acknowledgments}

\bibliographystyle{apsrev4-1}

\begin{thebibliography}{44}
\bibitem [{not()}]{note_general_HB}%
  \BibitemOpen
  \href@noop {} {}\bibinfo {note} {The general derivation shows that
  $\lambda=(3+n)/(6-2n)$ and $\mu=2n/(3-n)$, where $n$ is the HB exponent. In
  the case $n=1/2$, one has $\lambda=0.7$ and $\mu=0.4$ as used for rescaling
  the strain and stress in Fig.~\ref{fig:fluidity}(b). See companion paper for
  details \cite{companion}.}\BibitemShut {Stop}%
\end{thebibliography}

%



\clearpage
\newpage
\onecolumngrid
\setcounter{page}{1}
\setcounter{equation}{0}
\setcounter{figure}{0}
\global\def\thefigure{S\arabic{figure}}
\setcounter{table}{0}
\global\def\thetable{S\arabic{table}}

\begin{center}
    {\large\bf Stress Overshoots in Soft Glassy Materials}
\end{center}

\begin{center}
    {\large\bf {\sc Supplementary information}}
\end{center}

\section{Experimental details}

\seb{This section provides some details about the experiments used for the comparison with our fluidity model in the main text. Microgels were prepared from Carbopol ETD 2050 powder dispersed in water at weight concentrations $C$ ranging between 0.5\% and 3\%~wt. Carbopol ETD 2050 is made of homopolymers and copolymers of acrylic acid that are highly crosslinked with a polyalkenyl polyether. When neutralized using NaOH, the polymer particles swell and jam to form a dense, amorphous assembly of soft particles with typical size 6~$\mu$m \cite{Geraud:2017}. The reader is referred to Ref.~\cite{Divoux:2011} for the full preparation protocol.}

\begin{figure}[h!]
    \centering
   \includegraphics[width=0.6\columnwidth]{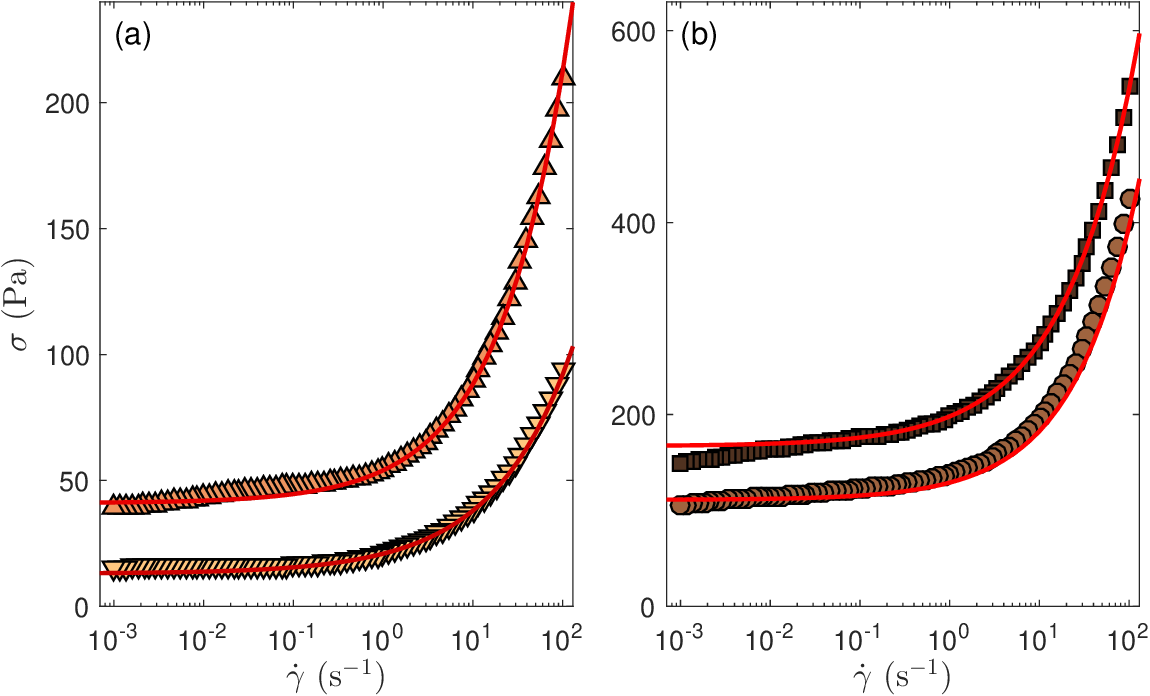}
    \caption{\seb{Flow curves, shear stress $\sigma$ vs shear rate $\gp$, of microgels made of Carbopol ETD~2050 with weight concentrations (a)~$C=0.5\%$~(\protect \invertedtriangleEXP) and 1\%~(\protect \triangleEXP), (b)~2\%~(\protect \circleEXP) and 3\%~wt~(\protect \squareEXP). The red solid lines show the Herschel-Bulkley behaviors, $\sigma=\sy+A\gp^n$, used for rescaling the data in the main text. The values of the yield stress $\sy$, the consistency index $A$, and the exponent $n$ are gathered in Table~\ref{table1}. Experiments performed at room temperature in a parallel-plate geometry of gap 1~mm and covered with sandpaper of roughness 46~$\mu$m. }}
    \label{fig:flowcurves}
\end{figure}

\begin{figure}[h!]
    \centering
   \includegraphics[width=0.9\columnwidth]{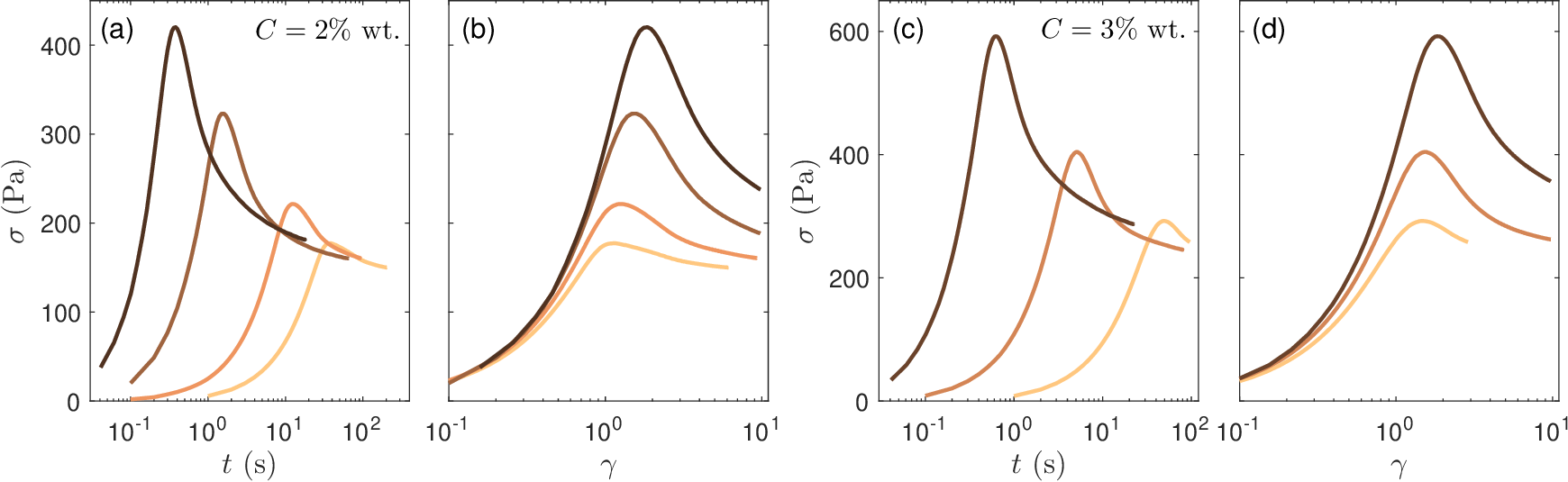}
    \caption{\seb{Stress responses recorded (a,b) in a 2~\% wt. Carbopol microgel under imposed shear rates $\gp=5$, 1, 0.2, and 0.03~s$^{-1}$ from left (darker color) to right (lighter color), plotted as a function of time $t$ in (a) and as a function of the strain $\gamma=\gp t$ in (b), and (c,d) in a 3~\% wt. Carbopol microgel under $\gp=3$, 0.3, and 0.03~s$^{-1}$ from left (darker color) to right (lighter color), plotted as a function of $t$ in (c) and as a function of $\gamma$ in (d). Experiments performed at room temperature in a parallel-plate geometry of gap 1~mm and covered with sandpaper of roughness 46~$\mu$m.}}
    \label{fig:suppovershoots}
\end{figure}

The experiments shown in Figs.~\ref{fig:overshoots}(a) and  \ref{fig:carbopol} in the main text were performed at room temperature in a parallel-plate geometry of radius 21~mm and gap 1~mm covered with sandpaper of roughness 46~$\mu$m. \seb{A stress-controlled rheometer (Anton Paar, MCR 301) imposes a constant shear rate $\gp$ to the sample thanks to a feedback loop on the shear stress $\sigma$. Figure~\ref{fig:flowcurves} shows the flow curves of the various samples together with the Herschel-Bulkley behaviors used for rescaling the data in the main text and summarized in Table~\ref{table1}. Additional stress responses showing some overshoots analyzed in Fig.~\ref{fig:carbopol} in the main text are shown in Fig.~\ref{fig:suppovershoots} for 2\% and 3\% wt. Carbopol microgels. More data can be found in Ref.~\cite{Divoux:2011} where the influence of boundary conditions was also explored.}

The experiments shown in Fig.~\ref{fig4} in the main text were performed at room temperature in a concentric-cylinder geometry of gap 1.1~mm covered with sandpaper of roughness 60~$\mu$m. \seb{The radius of the inner cylinder attached to the rotating shaft of the rheometer is 23.5~mm and the immersed height is 28~mm. Both cylinders are covered with sandpaper of roughness 60~$\mu$m. The velocity profiles presented in the lower inset in Fig.~\ref{fig4} were obtained through ultrasonic speckle velocimetry coupled to rheometry as introduced in Ref.~\cite{Manneville:2004a}. To provide acoustic contrast to the microgel and allow for local velocity measurements, hollow glass microspheres (Potters, Sphericel, mean diameter 6~$\mu$m density 1.1) were suspended within the initial Carbopol dispersion at a volume fraction of 0.5\%.}

\begin{table}[htb]
\begin{tabular}{c|c|c|c|c|c}
Symbol & $C$ ($\%$~wt) & $G_0$ (Pa) & $\sigma_\text{y}$ (Pa) & $n$ & $A$ (Pa.s$^n$) \\
\hline\hline
{$\triangledown$} & 0.5 & 72 $\pm 8$  & 13 $\pm 1$  & 0.50 $\pm 0.03$  & 7.9 $\pm 0.5$ \\   
{$\triangle$} & 1 & 142 $\pm 15$  & 41 $\pm 4$  & 0.56 $\pm 0.02$  & 13 $\pm 2$ \\   
{$\bullet$} & 2 & 285 $\pm 20$  & 111 $\pm 9$  & 0.60 $\pm 0.01$  & 18 $\pm 2$ \\
{$\square$} & 3 & 408 $\pm 30$  & 167 $\pm 20$  & 0.54 $\pm 0.02$  & 31 $\pm 3$  \\   

\end{tabular}
    \caption{Experimental parameters for the Carbopol microgels used in \seb{the present study}: weight concentration $C$, elastic modulus $G_0$, yield stress $\sigma_\text{y}$, shear-thinning exponent $n$ and the consistency index $A$ are inferred from Herschel-Bulkley fits of the steady-state \seb{flow curves, $\sigma$ vs $\dot\gamma$, shown in Fig.~\ref{fig:flowcurves}. The uncertainties reflect the typical variations of the various parameters due to sample volume variability from one loading to another and to possible solvent evaporation over long durations of a few hours, as well as the sensitivity of the HB parameters when fitting the flow curves over different ranges of shear     rates.}}\label{table1}
\end{table}

\end{document}